\documentstyle[12pt]{article}
\def\beq{\begin{equation}}   \def\eeq{\end{equation}}
\def\bea{\begin{eqnarray}}   \def\eea{\end{eqnarray}}

\newcommand{\matel}[3]{\langle #1|#2|#3\rangle}

\begin{document}

\begin{flushright}
UND-HEP-00-BIG\hspace*{.2em}06\\
%hep-ph/0009021\\
%{\footnotesize osakav2.tex}~~~~\today  \\ 
%version  2.0\\ 
\end{flushright}
\vspace{.3cm}
\begin{center} \Large 
{\bf Flavour Dynamics -- Central Mysteries of the 
Standard Model}
\footnote{Plenary Talk given at ICHEP2000, 
July 27 - August 2, 2000, Osaka, Japan}
\\
\end{center}
\vspace*{.3cm}
\begin{center} {\Large 
I. I. Bigi }\\ 
\vspace{.4cm}
{\normalsize 
{\it Physics Dept.,
Univ. of Notre Dame du
Lac, Notre Dame, IN 46556, U.S.A.} }
\\
\vspace{.3cm}
{\it e-mail address: bigi@undhep.hep.nd.edu } 
\vspace*{0.4cm}

{\Large{\bf Abstract}}
\end{center}

After pointing out the amazing success of the CKM description in 
accommodating the phenomenology of flavour changing neutral 
currents I review the
status of theoretical technologies  for extracting CKM parameters from
data. I sketch novel directions, namely
attempts to deal  with quark-hadron duality in a (semi)quantitative way and to
develop a QCD description of two-body modes of $B$ mesons. After commenting on
predictions  for $\epsilon ^{\prime}/\epsilon$ and CP asymmetries in $B$ 
decays I address indirect probes for New Physics in $D^0$ oscillations 
and CP violation, in $K_{\mu 3}$ decays and electric dipole 
moments. I describe in which way searching for New Physics 
in $B$ decays provides an exciting adventure with novel challenges 
not encountered before.

%\vspace*{.4cm}
%\vfill
%\noindent
%\vskip 5mm
%PACS 11.30.Er, 13.20.Eb, 13.25.Es
%\vskip 3mm

%]
%%%%%%%%%%%%
\tableofcontents 
%%%%%%%%%%

\vspace*{1.0cm}

Flavour dynamics involve central 
mysteries of the Standard Model (SM): Why is there a family 
structure relating quarks and leptons? Why is there more than 
one family, why three, is three a fundamental parameter? 
What is the origin of the observed pattern in the quark masses 
and the  
CKM parameters? 

There are two different strategies for obtaining answers to 
these questions: 

\noindent {\bf (A)} One argues that one has already enough data 
and therefore can turn one's energy towards the last fundamental 
challenge, namely to bring gravity into the quantum world; flavour 
dynamics with its family structure will then emerge as a 
`side product'. 

\noindent {\bf (B)} Suspecting that nature has a few more 
surprises up her sleeves one commits oneself to elicit more 
answers from her. 

My talk is geared towards strategy {\bf (B)} and its necessary 
theoretical tools. I will list experimental numbers without going into 
details; those can be found in Golutvin's talk \cite{GOLUTVIN}. 

%%%%%%%%%%%%%%%%%%%
\section{New Landmarks and Challenges}
%%%%%%%%%%%%%%%%%%%

Since ICHEP98 new landmarks have been  
reached:
\begin{itemize}
\item 
{\em Direct} CP violation has been established experimentally --  
a discovery of the first rank irrespective of 
theoretical interpretations. 
\item 
We are on the brink of observing CP violation in $B$ decays.
\item 
We are reaching fertile ground for finding New Physics in 
$D^0$ oscillations and CP violation.  

\end{itemize} 
On the theory side we are   
learning lessons of humility,  
increasing the sophistication of our theoretical technologies,  
and pushing back new frontiers.  

All of this leads to new challenges for theory, namely to regain 
theoretical control over $\epsilon ^{\prime}/\epsilon$; 
to develop 
reliable quantitative predictions for CP asymmetries 
in $B$ decays 
and to refine them into precise ones; to establish theoretical 
control over $D^0$ oscillations and CP violation 
and finally, to develop comprehensive strategies to not only 
establish the intervention of New Physics, but also identify its 
salient features. 

A major part of my talk will address extracting numerical values 
of CKM parameters; I will discuss  possible limitations to 
quark-hadron duality and refer to the 
lifetimes of charm and beauty hadrons as validation studies before 
describing new attempts to describe exclusive 
nonleptonic $B$ decays; I will sketch the difficulties inherent in 
predicting $\epsilon ^{\prime}/\epsilon$ before addressing 
CP violation in $B$ decays; I will comment on future searches 
for New Physics based on CKM trigonometry and the nature of theoretical 
uncertainties before describing `exotic' searches for transverse 
polarization of muons in $K_{\mu 3}$ decays, electric dipole moments 
and CP violation in charm transitions. 

%%%%%%%%%%%%%%%%%%%%%%%%%%%%%%
\section{The Charged Current Dynamics of Quarks} 
%%%%%%%%%%%%%%%%%%%%%%%%%
%%%%%%%%%%%%%%%%%%%%
\subsection{The `unreasonable' Success of the CKM Description}
%%%%%%%%%%%%%%%%%%%%%%%   
The observation of the `long' $B$ lifetime of about 1 psec together 
with the dominance of $b\to c$ over $b\to u$ revealed a hierarchical 
structure in the KM matrix that is expressed in the Wolfenstein 
representation in powers of $\lambda = {\rm tg}\theta _C$. We   
often see plots of the CKM unitarity triangle where the 
constraints coming from various observables appear as    
broad bands \cite{STOCCHI}. While the latter is often bemoaned, it obscures 
a more fundamental point: the fact that these constraints can be 
represented in such plots at all is quite amazing! Let me illustrate 
that by an analogy first: plotting the daily locations of the about 1000 
high energy physicists attending this meeting on a city map of Osaka 
produces fairly broad bands. Yet the remarkable thing is that these 
1000 people are in Osaka rather than spread over the world. On a map 
of Japan (let alone the world) these bands shrink to a point showing that 
the whereabouts of these phycisists follow an a priori 
highly unlikely distribution for which there must be a good reason. 
Likewise one should look at the 
bigger picture of flavour dynamics. 
The quark box 
{\em without} GIM subtraction yields a value for 
$\Delta m_K$ exceeding the experimental 
number by more than a factor of thousend; it is the GIM mechanism 
that brings it down to within a factor of two or so of experiment. 
The GIM subtracted quark box for $\Delta M_B$ coincides 
with the data again within a factor of two. Yet if the 
beauty lifetime were of order $10^{-14}$ sec while 
$m_t \sim 180$ GeV it would exceed it by 
an order of magnitude; on the other hand it would undershoot by an order 
of magnitude if $m_t \sim 40$ GeV were used with 
$\tau (B) \sim 10^{-12}$ sec; i.e., the observed value can be 
accommodated because a tiny value of $|V(td)V(ts)|$ is offset 
by a large $m_t$. 

This amazing success is repeated with $\epsilon$. Over the last 
25 years it could always be accommodated (apart from 
some very short periods of grumbling mostly off the record) 
whether the {\em correct} set [$m_t = 180$ GeV with $|V(td)|\sim 
\lambda ^3$, 
$|V(ts)|\sim \lambda ^2$] or the {\em wrong} one 
[$m_t = 40$ GeV with $|V(td)|\sim \lambda ^2$, 
$|V(ts)|=\lambda$] were used. Yet both 
$m_t = 180$ GeV with $|V(td)|=\lambda ^2$, 
$|V(ts)|=\lambda$ as well as 
$m_t = 40$ GeV with $|V(td)|=\lambda ^3$, 
$|V(ts)|=\lambda ^2$ would have lead to a clear inconsistency! 

Thus the phenomenological success of the CKM description has to be 
seen as highly nontrivial or `unreasonable'. This cannot have 
come about by accident -- there must be a good reason. 

%%%%%%%%%%%%%%%%
\subsection{Extracting CKM Parameters}
%%%%%%%%%%%%%%

A crucial element in extracting CKM parameters 
defined for the quark degrees
of freedom from data  involving hadrons is the quality of our theoretical
technologies to  deal with the strong forces. For strange mesons 
with $m_s < \Lambda _{QCD}$ one invokes chiral 
perturbation theory, for beauty hadrons with 
$m_b \gg \Lambda _{QCD}$ the heavy quark expansion 
(HQE) which might be extended to charm hadrons in a semiquantitative 
fashion ($m_c > \Lambda _{QCD}$)  
\footnote{The situation is qualitatively different for top states: 
with $\Gamma _t \sim {\cal O}(\Lambda _{QCD})$ top quarks decay 
before they can hadronize and they are therefore controlled by 
perturbative QCD \cite{DOK}. 
}. 
Lattice QCD on the other hand deals with the 
nonperturbative dynamics of all quark flavours.

Both HQE and lattice QCD (to be discussed in Kenway's lecture 
\cite{KENWAY})  
represent mature technologies with 
large common ground (both operate in Euklidean 
space) that are complementary to
each other. There has already been fruitful feedback between the two  
on the conceptual as well as numerical level; this interaction is about 
to intensify. While quark models no longer represent 
state-of-the-art, they still serve useful purposes in the diagnostics 
of our results if 
employed properly. 

The main tool for numerical results so far have been 
the HQE. The last few years have seen a conceptual covergence 
among its practitioners: most of them accept the argument 
that HQE allow to describe in principle nonleptonic as well as 
semileptonic beauty decays as long as an operator product 
expansion can be relied upon. At the same time one fully expects 
the numerical accuracy to decrease when going from 
$B \to l \nu q \bar q'$ to $B \to c \bar ud \bar q'$ and on to 
$B \to c \bar c s \bar q'$ for fundamental as well practical 
reasons (the latter meaning that the energy release is lowest for 
$b\to c \bar cs$.). Considerable progress has been achieved also 
in the numerical value of basic quantities the 
most important one being the beauty quark mass. Last year three 
groups extending earlier work by Voloshin \cite{VOL} 
have presented new 
extractions from data, which -- when expressed in terms of the 
socalled `kinetic' mass -- read as follows: 
\beq 
m_b^{\rm kin} (1\, {\rm GeV}) = 
4.56 \pm 0.06  \; \; 
{\rm GeV} \; \;  \cite{MEL}, \; 
4.57 \pm 0.04  \; \; 
{\rm GeV} \; \;  \cite{HOANG}, \; 
4.59 \pm 0.06  \; \; 
{\rm GeV} \; \;  \cite{SIGNER}
\eeq
The error estimates of 1 - 1.5 \% might be overly optimistic (as it 
often happens), but not foolish. Since all 
three analyses use basically the same input from the 
$\Upsilon (4S)$ region, they could suffer from a common 
systematic uncertainty, though. This can be checked by analysing 
the shape of the lepton spectrum in $B \to l \nu X$. More 
concretely one forms two moments both of the lepton and of the hadron 
energies \cite{MOM}; each set yields $\bar \Lambda$ and 
$\mu _{\pi}^2$, 
where $\bar \Lambda \to M_B - m_b$ as $m_b \to \infty$ and 
$\mu _{\pi}^2 \equiv \matel{B}{\bar b (iD)^2b}{B}/2M_B$. 
Comparing those two sets of values with each other 
and with the $m_b$ values listed above represents 
a crucial self-consistency check. An  early CLEO analysis 
appeared to yield inconsistent values. It is being redone 
now, and I eagerly await their results; yet I do that 
with considerable confidence, in particular since a recent 
lattice study \cite{SIMONE} has yielded numbers that are 
in agreement with those inferred from the SV sum rules 
\cite{OPTICAL}. 

Two methods exist with excellent theoretical credentials for 
determining $V(cb)$: 

(i) Extrapolating the rate of $B \to l \nu D^*$ to 
zero recoil one extracts $V(cb)F_{D^*}(0)$. The form factor 
$F_{D^*}(0)$ has the nice features that it is normalized 
to unity in the infinite mass limit and that the leading 
nonperturbative correction is of order $1/m_Q^2$. 
Unfortunately it is $m_c$ that sets the scale here rather 
than $m_b$, and that is one of the 
challenges in evaluating it. Three estimates provide representative 
numbers: 
\beq 
F_{D^*}(0) = 0.89 \pm 0.08  \; \;       \cite{URI1}, \; 
0.913 \pm 0.042 \; \; \cite{BABARBOOK}, \; 
0.935 \pm 0.03 \; \;  \cite{LAT}  
\eeq 
I will use here 
\beq 
F_{D^*}(0) = 0.90 \pm 0.05 
\label{FD}
\eeq 
as a convenient {\em reference} point. 
CLEO has presented a new analysis that yields 
a considerably larger number than before \cite{ECKLUND}:  
\beq
|V(cb)F_{D^*}(0)|  = (42.4 \pm 1.8|_{stat} \pm 
1.9|_{syst} )\times 10^{-3}  
\eeq  
The updated LEP number on the other hand has hardly changed 
\cite{BARBERIO}:  
\beq
|V(cb)F_{D^*}(0)|  = (34.9 \pm 0.7|_{stat} \pm 
1.6|_{syst} )\times 10^{-3} 
\eeq
There is now about a 20\% difference between the two 
central values, which means that `stuff happens'. With 
Eq.(\ref{FD}) one gets:  
\bea 
|V(cb)|_{excl,CLEO}&=& (47.1 \pm 2.0|_{stat} \pm 
2.1|_{syst} \pm 2.1_{th} )\times 10^{-3}\\
|V(cb)|_{excl,LEP}&=& (38.8 \pm 0.8|_{stat} \pm 
1.8|_{syst} \pm 1.7_{th} )\times 10^{-3}
\eea  
I view a theoretical error of 5\% as on the optimistic side, 
and I am skeptical about being able to reduce it below this 
level.  

(ii) The {\em inclusive} semileptonic width of $B$ mesons 
can be calculated 
in the HQE: $\Gamma _{SL}(B) \propto m_b^5 \cdot 
(1+ {\cal O}(1/m^2_b) + {\cal O}(\alpha _S))$. Again there is no 
correction $\sim {\cal O}(1/m_Q)$. The advantage over the 
previous case is that the expansion parameter is effectively 
the inverse energy release $\sim (m_b - m_c)^{-1}$ rather than the 
larger $1/m_c$; the challenge is provided by 
the fact that the leading term depends on the fifth power of the 
$b$ {\em quark} mass. It was only the great conceptual and technical 
progress in HQE that made this method competitive.  

LEP has updated its analysis and finds: 
\beq 
|V(cb)|_{incl} = (40.76 \pm 0.41|_{stat} \pm 2.0_{th} )\times 10^{-3}
\eeq  
The theoretical error has been evaluated in a fairly careful way 
\cite{HQR}; 
I am quite optimistic that it can be cut in half 
in the foreseeable future; but even then it would appear to 
represent the limiting factor. Yet it is mandatory to check the small 
overall experimental error. CLEO has amassed a huge amount of data 
on tape; I am most eager to see their findings.  
 
The first {\em direct} evidence for $V(ub) \neq 0$ came from the 
endpoint spectrum in {\em inclusive} semileptonic $B$ decays. 
Such studies yielded $|V(ub)|_{end} = (3.2 \pm 0.8)\times 10^{-3}$ 
with a heavy reliance on theoretical models which makes 
both the central value and the error 
estimate suspect. Yet with huge new data sets becoming available, 
this avenue should be re-visited due to the following two observations: 
\begin{itemize}
\item 
The $AC^2M^2$ model constitutes a good implementation of QCD, in 
particular for $b\to u$ transitions \cite{ACCMM}. The main caveat is that 
one should not determine the two model parameters $p_F$ and 
$m_{sp}$ from the $b\to c$ spectrum and then apply it 
blindly to $b\to u$ decays. With sufficient statistics one can 
fit it directly to the $b\to u$ spectrum even over 
the very limited kinematical regime where it can be cleanly separated 
from $b\to c$. 
\item 
A few years ago it has been suggested \cite{MOTION} to extract the  
required shape function for $b\to u$ from the measured   
photon spectrum in $B \to \gamma X$. This might become a feasible 
procedure with future data. Some more theoretical work 
is needed, though, a point I will return to.  

\end{itemize} 
From the exclusive channels $B\to l \nu \pi$ and 
$B \to l \nu \rho$ one has inferred 
\beq 
|V(ub)|_{excl} = (3.25 \pm 0.14|_{stat.} \pm 0.27|_{syst.} 
\pm 0.55|_{th}) \times 10^{-3} 
\eeq 
There is a very strong model dependance, and it is 
quite  unclear to 
me whether the theoretical uncertainty has been evaluated in a 
reliable fashion by comparing the findings from various quark 
models and QCD sum rules. One hopes that lattice QCD will 
provide the next step forward. 

LEP groups have made the heroic effort to extract the total 
width $\Gamma (H_b \to l \nu X_{no \; charm})$. Their findings 
read as follows \cite{GOLUTVIN}:
\beq 
|V(ub)|_{\Gamma _{SL}} = (4.04 \pm 0.44|_{stat} 
\pm 0.46|_{b\to c,syst} \pm 0.25|_{b\to u,syst} 
\pm 0.02 |_{\tau _b} \pm 0.19 |_{HQE} ) \times  10^{-3} 
\eeq
The theoretical uncertainties in this fully integrated 
width are under good control \cite{GSLBU}; however it is 
an experimental tour de force, as already indicated by the 
errors, with the uncertainty in the modelling for 
$b \to c$ the central one.

The main drawback in using the charged lepton energy as a 
kinematical discriminator is its low efficiency: about 90\% 
of the $b\to u$ events are buried under the huge 
$b\to c$ background.  The hadronic recoil mass spectrum 
$\frac{d}{dM_X}\Gamma (B\to l \nu X)$ provides a much more 
efficient filter with only about 10\% of $b\to u$ being swamped 
by $b\to c$ as first suggested within a parton model description 
\cite{KIM}. Using HQE methodology it has been shown that the 
theoretical description can be based more directly on 
QCD \cite{DIKE,FALIWI}. Furthermore the fraction of 
$b\to u$ events below $M_X \sim 1.6$ GeV appears to be 
fairly stable. The predicted $M_X$ spectrum can be compared 
with data -- if one `smears' the latter over energy intervals 
$\sim \Lambda$. Refinements of these ideas are under active 
theoretical study \cite{BAUER}. 

(i) $|V(td)|$ can be inferred from $B_s$ oscillations 
\footnote{The 3-family unitarity constraint 
$|V(ts)| \simeq |V(cb)|$ is assumed throughout this talk unless 
stated otherwise.} 
\beq 
\frac{x_d}{x_s} \simeq \frac{|V(td)|^2}{|V(ts)|^2} 
\frac{Bf^2(B_d)}{Bf^2(B_s)} \; , 
\eeq 
although even the relative size of $B_d$ and $B_s$ oscillations 
could be affected significantly by New Physics. (ii) Another approach is 
to compare {\em exclusive} radiative decays 
$B\to \gamma \rho /\omega$ vs. $B \to \gamma K^*$. Yet one has to keep 
in mind here that long distance physics could affect 
$B\to \gamma \rho$ much more than $B\to \gamma K^*$. 
(iii) The cleanest way theoretically is provided by the width for 
$K^+ \to \pi ^+ \nu \bar \nu$. With the hadronic matrix element 
inferred from $\Gamma (K^+ \to \pi ^0 l^+ \nu)$ the contributions from 
intermediate charm quarks provide the irreducible theoretical 
uncertainty estimated to be around several percent. 
With the present loose bounds on $|V(td)|$ one expects 
\cite{BUCHBURAS} 
\beq 
{\rm BR} (K^+ \to \pi ^+ \nu \bar \nu ) = (0.82 \pm 0.32) 
\cdot 10^{-10} 
\eeq
One candidate has been observed by E787 at BNL corresponding to 
\beq 
{\rm BR} (K^+ \to \pi ^+ \nu \bar \nu ) = (1.5 ^{+3.4}_{-1.2} ) 
\cdot 10^{-10} 
\eeq
The single event sensitivity is supposed to go down to 
$0.7 \cdot 10^{-10}$; the successor experiment E949 hopes for 
a sensitivity of $\sim 10^{-11}$. 

{\em In summary:} 
\begin{itemize}
\item 
There are two ways for extracting $|V(cb)|$ from semileptonic 
$B$ decays where the {\em theoretical} uncertainty has been reduced 
to about 5\% with a further reduction appearing feasible. 
This theoretical confidence cannot be put to the test yet due to a 
divergence in the available data. 
\item 
PDG2K quotes a $\sim 40$ \% error on $V(ub)$. 
The situation will improve qualitatively as well as 
quantitatively: reducing uncertainties down to the 
10\% level seems feasible, and in the long run one can dream 
to go even beyond that! 
\item 
Observing $B_s$ oscillations and $B \to \gamma \rho/\omega$ 
would elevate our knowledge of $|V(td)|$ to a new level: 
in particular the former should yield a value with an error not 
exceeding 10 \%, although it could be affected very significantly 
by New Physics; an intriguing long term prospect is provided by 
$K^+ \to \pi ^+ \nu \bar \nu $. 

\end{itemize}

%%%%%%%%%%%%%%%%%%
%\section{Nonleptonic Heavy Flavour Decays}
%%%%%%%%%%%%%%%%
%%%%%%%%%%%%%%%%%%%
\subsection{Quark-Hadron Duality -- a New Frontier}
%%%%%%%%%%%%%%%%%

When extracting the value of CKM parameters with few 
percent errors only, one has to be concerned about several sources 
of {\em systematic} uncertainties, prominent among them 
theoretical ones. A fundamental one  
is the assumption of {\em quark-hadron duality (QHDu)} that enters 
at various stages of the theoretical reasoning. When calculating 
a rate on the quark-gluon level QHDu is invoked to equate the result 
with what one should get for the corresponding process expressed in
hadronic quantities. 

QHDu {\em cannot} be exact: it is an approximation the quality of 
which is process-dependant -- it should work better for semileptonic 
than nonleptonic transitions -- and increases with the amount of 
averaging or `smearing' over hadronic channels. There is a lot of 
folklore that leads to several useful concepts -- but no theory. 
That is not surprising: for QHDu can be addressed in a 
quantitative fashion only {\em after} nonperturbative 
effects have been brought under control, and that has happened only 
relatively recently in beauty decays. 

Developing such a theory for QHDu thus represents a 
new frontier requiring the use of new tools.  
Considerable insight exists into the physical origins of QHDu 
violations: (i) They are caused by the exact location of 
hadronic thresholds that are notoriously hard to evaluate. 
Such effects are implemented through `oscillating terms'; i.e.,  
the fact that innocuous, since suppressed contributions 
${\rm exp}(-m_Q/\Lambda )$ in Euclidean space turn into 
dangerous while unsuppressed sin$(m_Q/\Lambda )$ terms in 
Minkowski space. (ii) There is bound to be some sensitivity 
to `distant cuts' \cite{OPTICAL}. (iii) The validity of the 
$1/m_c$ expansion arising in the description of 
$B\to l \nu D^*$ is far from guaranteed. 

The OPE {\em per se} is insensitive to QHDu violations (although it 
provides some indirect qualitative insights). One can probe 
QHDu in exactly solvable model field theories among which the 
't Hooft model -- QCD in 1+1 dimensions with 
$N_C \to \infty$ -- has gained significant consideration. 
It had been suggested \cite{GL}, based on a numerical analysis, that 
nonleptonic transitions exhibit significant or even large 
QHDu violations; yet analytical studies revealed such violations to be 
tiny only \cite{THOOFT}, even in spectra \cite{LU}. 

A more convincing probe for QHDu violations would be based on a 
procedure familiar from experimental analyses: one employs different 
methods to determine the same basic quantity. I have already 
listed one example, namely to extract $m_b$ from 
$\Upsilon (4S)$ spectroscopy as well as the leptonic and 
hadronic moments in $B$ decays. One very telling implementation 
of such a program would be to determine CKM parameters in 
$B_s$ decays and compare the results with the findings 
in $B_{u,d}$ decays. For practical reasons
one would probably be limited to compare the 
leptonic and hadronic moments in semileptonic $B_s$ decays 
and to infer $|V(bc)|$ from $\Gamma _{SL}(B_s)$ and 
$B_s \to l \nu D_s^*$. Comparing $B_s$ with $B$ results is 
much more revealing than comparing $B_d$ with $B_u$ decays. 
For a likely source of QHDu violations in $b \to c$ 
is provided by the 
presence of a `near-by' resonance with appropriate quantum numbers. 
If $B_d$ decays are affected, so will be those of $B_u$ and by the same 
amount, but not $B_s$. Likewise a resonance near the $B_s$ 
could affect its transitions, but not those of $B_{u,d}$. Nature
had to be truly malicious to place one resonance next to 
$B_{u,d}$ and a second one 
next to the
$B_s$. Barring that a comparison of the values  of $|V(cb)|$ 
obtained from $B$
and
$B_s$ decays would allow us to gauge  QHDu violations in those 
transitions. The situation is more complex for $b\to u$, though, as 
already alluded to. 
An isoscalar or isovector resonance would affect $B_u$ and $B_d$ 
modes differently.   
 
%%%%%%%%%%%%%%%%%%%%%%%%%%
\subsection{Lifetimes as Validation Studies}
%%%%%%%%%%%%%%%%%%%%%%%

Among the many several important lessons to be derived 
from the lifetimes of charm
and  beauty hadrons I will emphasize just one aspect: with QHDu
violations  expected to be larger in nonleptonic than semileptonic decays, one
can  view studies of lifetimes as validation studies. The new measurements
reported on 
$D^0$, $D^+$, $D_s$ and $\Lambda _c$ are in line with previous 
mesasurements and do not change the overall picture 
\cite{BELLINI}: 
(i) The 
$D^0$-$D^+$ lifetime difference is given mainly by 
Pauli intereference 
yielding a ratio of $\sim 2 \cdot (f_D/200\, {\rm MeV})^2$. 
(ii) Weak
annihilation should contribute in  {\em mesons} on the 10 - 20 \% level. 
(iii)
The ratio 
$\tau (D_s)/\tau (D^0)$ is fully consistent with such a 
semiquantitative picture. (iv) What is missing for a 
full evaluation are more accurate $\Xi _c$ lifetimes: 
measurements of $\tau (\Xi _c^{0,+})$ with 10 - 15 \% accuracy 
are needed for this purpose. 

The situation of beauty lifetimes has changed in one respect: 
the world average for the $B^+$-$B_d$ lifetime ratio now shows 
a significant excess over unity in agreement with a prediction 
using factorization: 
\beq 
\frac{\tau (B^-)}{\tau (B_d)} = 1.07 \pm 0.02 \; \; {\rm exp. \; 
\cite{GOLUTVIN}} 
\; \; \; vs. \; \; \; 1 + 0.05 \cdot \left( 
\frac{f_B}{200 \; {\rm MeV}}\right) ^2 \; \;  {\rm theor. \; 
\cite{BELLINI}} 
\eeq
The discrepancy for $\tau (\Lambda _b)$ has remained basically the 
same:  
\beq 
\frac{\tau (\Lambda _b)}{\tau (B_d)} = 0.794 \pm 0.053 \; \; {\rm exp. 
\; \cite{GOLUTVIN}} 
\; \; \; vs. \; \; \; 0.88 - 1.0 \; \;  {\rm theor. \; 
\cite{BELLINI} } 
\eeq
While this could signal a significant limitation to QHDu, I like to 
reserve my judgement till CDF and D0 measure 
$\tau (\Lambda _b)$ \& $\tau (\Xi _b^{0,-})$ in the next run. 

The most striking success has been the apparently correct prediction 
of the $B_c$ lifetime: $\tau (B_c) \sim 0.5$ psec \cite{MSTM} vs. 
the CDF findings 
$0.46 \pm 0.17$ psec with $\tau (B_c)/\tau (B_d) \sim 1/3$: the 
absence of a $1/m_Q$ correction is essential here. The 
$B_s$ lifetime deserves further dedicated scrutiny since 
theoretically one expects with confidence 
$\bar \tau (B_s)/\tau (B_d) = 1 \pm {\cal O}(0.01)$ vs. the 
experimental value of $0.945 \pm 0.039$. 

%%%%%%%%%%%%%%%%
\subsection{Exclusive Nonleptonic $B$ Decays -- another 
New Frontier}
%%%%%%%%%%%%

In describing nonleptonic two-body modes $B\to M_1 M_2$ 
valuable guidance has been provided by symmetry considerations 
based on $SU(2)$ and to a lesser degree $SU(3)$. 
Phenomenological models have played an important role; more 
often than not they involve factorization as a central 
assumption. Such models still play an important role in 
widening our horizon when used with common sense \cite{FLETAL}. 
Yet the bar has been raised for them by the emergence of a new 
theoretical framework for dealing with these decays. The 
essential pre-condition for this framwork is the large energy 
release, and it invokes concepts like `colour transparency' 
\cite{WARD}; while 
those have been around for a while, only now they 
are put into a 
comprehensive framework. Two groups have presented results on 
this \cite{BBNS,KLS}. The common feature in their approaches is that 
the decay amplitude is described by a kernel containing the 
`hard' interaction given by a perturbatively evaluated effective 
Hamiltonion folded with form factors, decay constants and ligh-cone 
distributions into which the long distance effects are lumped; this 
{\em factorization} is symbolically denoted by 
\beq 
\matel{M_1M_2}{ H}{B} = f_{B\to M_1}f_{M_2}T^H * \Phi _{M_2} + ... 
\eeq  
The two groups differ in their dealings with the soft part: 
BBNS regularize the divergent IR integrals they encounter at the 
price of introducing low energy parameters. KLS on the other hand 
invoke Sudakov form factors to shield them against IR singularities. 
It is not surprising that the two groups arrive at different 
conclusions: while BBNS infer final state interactions to be mostly 
small in $B\to \pi \pi, K \pi$ with weak annihilation being suppressed, 
KLS argue for weak annihilation to be important with final state 
interactions {\em not} always small. 

The trend of these results have certainly the ring of truth for me: 
e.g., while factorization represents the leading effect in most cases 
(including $B\to D\pi$), it is not of universal quality. One should 
also note that the {\em non}-factorizable contributions move the 
predictions for branching ratios towards the data -- a feature one could 
not count on {\em a priori}. It is not clear to me yet whether the 
two approaches are complementary or irreconcilable. Secondly one should view
these  predictions as preliminary: a clear disagreement with future data 
should be taken as an opportunity for learning rather than 
for discarding the 
whole approach. This is connected with a third point: there are 
corrections of order $\Lambda /m_b$ which are beyond our 
computational powers. Since $\Lambda$ might be as large as 0.5 - 1 GeV, 
they could be sizeable.

%%%%%%%%%%%%%%%%%%%%%%%%
\subsection{Radiative $B$ Decays}
%%%%%%%%%%%%%%

The transition $B\to \gamma X$ has been the first correctly 
predicted penguin footprint. The CLEO number is still the most accurate 
one, but the BELLE result is not far behind 
\bea 
{\rm BR}(B\to \gamma X_{no \; charm}) &=& 
(3.15 \pm 0.35 \pm 0.32 \pm 0.26) \cdot 10^{-4} \; \; {\rm CLEO}\\
{\rm BR}(B\to \gamma X_{no \; charm}) &=& 
(3.34 \pm 0.5 \pm 0.35 \pm 0.28) \cdot 10^{-4} \; \; {\rm BELLE}
\eea 
The SM prediction as summarized in an illuminating talk by Misiak 
reads \cite{MISIAK}  
\beq 
{\rm BR}(B\to \gamma X_{no \; charm})|_{\rm SM} =  
(3.29 \pm 0.33) \cdot 10^{-4}
\eeq 
While the central value and the uncertainty have hardly 
changed over the last four years, an impressive theoretical 
machinery has been developed resulting in many new calculations -- 
with the result that new contributions largely cancel. 
Careful analysis of the photon spectrum is under way, which is necessary 
to determine the branching ratio even more precisely and to determine 
the shape function needed to extract $|V(ub)|$ from the lepton endpoint 
spectrum \cite{MOTION}. 

The results and caveats for $B\to l^+l^- X$ have been updated. One 
should note that New Physics in general impacts $B\to \gamma X$ and 
$B \to l^+l^- X$ quite differently. 

%%%%%%%%%%%%%%%%%%%%
\section{CP Violation in $\Delta S, \Delta B \neq 0$}
%%%%%%%%%%%%%%%

The quantity $\epsilon ^{\prime}/\epsilon$ describes the difference in 
CP violation between  
$K_L \to \pi ^+\pi^-$ and $K_L \to \pi ^0\pi^0$: 
\beq 
{\rm Re} \frac{\epsilon ^{\prime}}{\epsilon } = 
\frac{1}{6} \left[ 
\frac{\Gamma (K_L\to \pi ^+\pi^-)/\Gamma (K_S\to \pi ^+\pi^-)}
{\Gamma (K_L\to \pi ^0\pi^0)/\Gamma (K_S\to \pi ^0\pi^0)} -1 \right] 
\eeq
Within the KM ansatz direct CP violation has to exist, yet it is suppressed 
by the $\Delta I=1/2$ rule and the large 
top mass: $0< \epsilon ^{\prime}/\epsilon \ll 1/20$. A  
guesstimate suggests  
$\epsilon ^{\prime}/\epsilon \sim {\cal O}(10^{-3})$ \cite{FABB}. The 
effective CP odd $\Delta S=1$ Lagrangian has been calculated with 
high accuracy on the quark level \cite{BURAS}; eight operators emerge. 
Evaluating their hadronic matrix elements with the available techniques 
one finds four positive and four negative contributions of roughly 
comparable size giving rise to large cancellations and thus 
enhanced uncertainties with central values typically below 
$10^{-3}$. While such studies found sizeable $\Delta I=1/2$ 
enhancements they fell well short of the observed size; 
various rationalizations were given for this failure, and 
overcoming it was left as a homework assignment for lattice QCD. 
However there were dissenting voices arguing for a more phenomenological 
approach where reproducing the $\Delta I=1/2$ 
rule is imposed as a goal. Not surprisingly this required the enhancement 
of some operators more than others thus reducing the aforementioned 
cancellations and increasing the prediction for 
$\epsilon ^{\prime}/\epsilon$ \cite{TRIESTE}. The first KTeV data gave
considerable respectability to this approach and lead to 
re-evaluations of
other studies leading to somewhat larger  predictions, as discussed at this
conference \cite{VALENCIAETAL}. 

This illustrates that theoretical uncertainties are very hard 
to estimate reliably, although in fairness two things should 
be pointed out: (i) Due to the large number on contributions with 
different signs one is facing an unusually complex situation. 
(ii) While there is no doubt that $\epsilon ^{\prime}\neq 0$ 
holds, its exact size is still uncertain: 
\beq 
{\rm Re}\left( \frac{\epsilon ^{\prime}}{\epsilon ^{\prime}}
\right) = 
(2.80 \pm 0.41)\cdot 10^{-3} \; \; \; {\rm KTeV}, \;  
(1.40 \pm 0.43)\cdot 10^{-3} \; \; \; {\rm NA48} \; ; 
\eeq
some of the earlier 
theoretical expectations might experience some vindication still. 
In any case we are eagerly awaiting the new results from KTeV. 

Our interpretation of the data is thus still in limbo: 
it might represent another striking success for the KM scheme with 
the $\Delta I=1/2$ rule explained in one fell swoop -- or it might 
be dominated by New Physics. I am not very confident that analytical 
methods can decide this issue, although some interesting 
new angles have been put forward on the $\Delta S=1/2$ 
rule \cite{PRADES}. One has to hope for
lattice QCD  to come through, yet it has to go beyond the 
quenched approximation, which will require more time. 

Although CP violation implies T violation due to the CPT theorem, 
I consider it
highly significant that more  direct evidence has been obtained through the 
`Kabir test': CPLEAR has found \cite{CPLEAR} 
\beq 
A_T \equiv 
\frac{\Gamma (K^0 \to \bar K^0) - \Gamma (\bar K^0 \to K^0)}
{\Gamma (K^0 \to \bar K^0) + \Gamma (\bar K^0 \to K^0)} = 
(6.6 \pm 1.3 \pm 1.0)\cdot 10^{-3} 
\eeq
versus the value $(6.54 \pm 0.24)\cdot 10^{-3}$ inferred from 
$K_L \to \pi^+\pi^-$. Of course, some assumptions still 
have to be made, namely that {\em semileptonic} $K$ decays obey 
CPT or that the Bell-Steinberger relation is satisfied with 
{\em known} decay channels only. Avoiding both assumptions 
one can write down an 
admittedly contrived scheme where the CPLEAR data are  
reproduced {\em without} T violation; the price one pays is a large CPT 
asymmetry $\sim {\cal O}(10^{-3})$ in 
$K^{\pm} \to \pi ^{\pm}\pi ^0$ \cite{TBS}. 

KTeV and NA48 have analyzed the rare decay 
$K_L \to \pi^+\pi^- e^+e^-$ and found a large {\em T-odd} 
correlation between the $\pi^+\pi^-$ and $e^+e^-$ planes in 
full agreement with predictions \cite{SEHGAL}. 
Let me add just two comments here: (i) This agreement cannot be 
seen as a success for the KM ansatz. Any scheme reproducing  
$\eta _{+-}$ will do the same. (ii) The argument that strong final state 
interactions (which are needed to generate a T odd correlation 
above 1\% with T invariant dynamics) cannot affect the 
relative orientation 
of the $e^+e^-$ and $\pi ^+\pi ^-$ planes fails on the 
quantum level \cite{TBS}.

%%%%%%%%%%%%%%%%%%%%
%\section{CP Violation in $\Delta B\neq 0$}
%%%%%%%%%%%%%

One often hears that observing a CP asymmetry in 
$B\to \psi K_S$ is no big deal since it is confidently expected -- 
unless it clearly falls outside the predicted range -- and likewise 
in $B\to \pi^+\pi^-$ since it cannot be interpreted cleanly due to 
Penguin `pollution' and the value of its asymmetry is hardly constrained. 
Such sentiments, however, miss the paradigmatic character of such 
observations: (a) An asymmetry in $B\to \psi K_S$ would be the first 
one observed outside $K_L$ decays, it would have to be big to be 
established in the near future and it would establish the KM ansatz 
as a major agent. (b) Likewise an asymmetry in $B\to \pi^+\pi^-$ 
again would have to be big, and it would probably reveal 
{\em direct} CP violation to be big as well in beauty decays. 

These CP asymmetries are described in terms of the angles of the 
usual unitarity triangle. An ecumenical message in PDG2000 
endorses two different notations, namely 
\beq 
\phi _1 \equiv \beta = \pi - 
{\rm arg}\left( \frac{V_{tb}^*V_{td}}{V_{cb}^*V_{cd}}
\right) ,  
\phi _2\equiv \alpha =  
{\rm arg}\left( \frac{V_{tb}^*V_{td}}{-V_{ub}^*V_{ud}} 
\right) ,  
\phi _3 \equiv \gamma =  
{\rm arg}\left( \frac{V_{ub}^*V_{ud}}{-V_{cb}^*V_{cd}}\right) . 
\eeq
From CP insensitive rates one can deduce the sides of this triangle 
and from CP asymmetries the angles: e.g., from 
$\epsilon/\Delta m(B_d)$ one can infer sin$2\phi_1$. A whole 
new industry has sprung up for doing these fits. Typical 
examples are (I will discuss caveats below): 
\bea 
{\rm sin}2\phi _1 &=& 0.716 \pm 0.070  \; \cite{STOCCHI} \; 
\leftrightarrow \; 0.7 \pm 0.1  \; \cite{COMBINER} \\
{\rm sin}2\phi _2 &=& -0.26 \pm 0.28  \; \cite{STOCCHI} \; 
\leftrightarrow \; - 0.25 \pm 0.6  \; \cite{COMBINER}   
\eea
The first results from the asymmetric $B$ factories leave us 
in limbo:
\bea 
{\rm sin}2\phi _1 &=& 0.45 ^{+0.43 + 0.07}_{-0.44 - 0.09}  \; \; 
{\rm BELLE} \\
{\rm sin}2\beta &=& 0.12 \pm 0.37 \pm 0.09  \; \; 
{\rm BaBar}
\eea 
Nevertheless one can raise the question what we would learn 
from a `Michelson-Morley outcome', if, say, 
$|{\rm sin}2\phi_1| < 0.1$ were established? Firstly, we would know 
that the KM ansatz would be ruled out as a major player 
in $K_L \to \pi \pi$ -- there would be no plausible deniability! 
Secondly, one would have to raise the basic question why the 
CKM phase is so suppressed, unless there is a finely tuned cancellation 
between KM and New Physics forces in $B\to \psi K_S$; this would shift 
then the CP asymmetry in $B\to \pi \pi, \, \pi \rho$.

%%%%%%%%%%%%%%%%
\section{Probing for New Physics}
%%%%%%%%%%%%%%

$\Delta S=1,2$ dynamics have provided several examples of revealing the 
intervention of features that represented New Physics 
{\em at that time}; 
it thus has been instrumental in the evolution of the SM. This happened 
through the observation of `qualitative' discrepencies; i.e., 
rates that were expected to vanish did not, or rates were 
found to be smaller than expected by several orders of magnitude. 
Such an indirect search for New Physics can be characterised as a 
`King Kong' scenario: one might be unlikely to encounter King 
Kong; yet once it happens there can be no doubt that one has 
come across someting out of the ordinary. Such a situation 
can be realized for charm and $K_{\mu 3}$ decays and EDMs.

%%%%%%%%%%%%%
\subsection{$D^0$ Oscillations \& CP Violation}
%%%%%%%%%%%%%%%

It is often stated that $D^0$ oscillations are slow and 
CP asymmetries tiny within the SM and that therefore their analysis 
provides us with zero-background searches for New Physics. 

Oscillations are described by the normalized mass and width 
differences:  
$x_D \equiv \frac{\Delta M_D}{\Gamma _D}$,   
$y_D \equiv \frac{\Delta \Gamma}{2\Gamma _D}$.  
A conservative SM estimate yields $x_D$, $y_D$ $\sim 
{\cal O}(0.01)$. Stronger bounds have appeared in the literature, 
namely that the OPE contributions are completely insignificant 
and that long distance contributions {\em beyond} the OPE provide the 
dominant effects yielding $x_D^{SM}$, $y_D^{SM}$ 
$\sim {\cal O}(10^{-4} - 10^{-3})$. A recent detailed analysis 
\cite{BUOSC} revealed 
that a proper OPE treatment reproduces also such long distance 
contributions with  
\beq 
x_D^{SM}|_{OPE}, \, y_D^{SM}|_{OPE} \sim {\cal O}(10^{-3}) 
\eeq   
and that $\Delta \Gamma $, which is generated from  
on-shell contributions, is -- in contrast to $\Delta m_D$
-- insensitive to New Physics while on the other hand more susceptible 
to violations of QHDu. 

Four experiments have reported new data on $y_D$ \cite{GOLUTVIN}: 
\bea
 y_D = (0.8 \pm 2.9 \pm 1.0) \% \; \; {\rm E791}&,& \; 
(3.42 \pm 1.39 \pm 0.74) \% \; \; {\rm FOCUS} \\
y_D = (1.0^{+3.8 + 1.1}_{-3.5-2.1})\% \; \; {\rm BELLE} &,& \; 
y_D^{\prime} = (-2.5 ^{+1.4}_{-1.6} \pm 0.3)\% \; \; {\rm CLEO}
\eea
E 791 and FOCUS compare the lifetimes for two different channels, 
whereas CLEO fits a general lifetime evolution to 
$D^0(t) \to K^+\pi ^-$; its $y_D^{\prime}$ depends on the strong 
rescattering phase between $D^0 \to K^-\pi^+$ and 
$D^0 \to K^+\pi^-$ and therefore could differ substantially from 
$y_D$ -- even in sign \cite{PETROV} -- if that phase were 
sufficiently large. 
The FOCUS data contain a suggestion that the lifetime 
difference in the 
$D^0 - \bar D^0$ complex might be as large as ${\cal O}(1\% )$. 
{\em If} $y_D$ indeed were $\sim 0.01$, two scenarios could arise 
for the mass difference. If $x_D \leq {\rm few} \times 10^{-3}$ 
were found, one would infer that the $1/m_c$ expansion yields a 
correct semiquantitative result while blaming the large value for 
$y_D$ on a sizeable and not totally surprising violation of 
QHDu. If on the other hand $x_D \sim 0.01$ would emerge, we would face 
a theoretical conundrum: an interpretation ascribing this to New 
Physics would hardly be convincing since $x_D \sim y_D$. A more sober 
interpretation would be to blame it on QHDu violation or on the 
$1/m_c$ expansion being numerically unreliable. Observing 
$D^0$ oscillations then would not constitute a `King Kong' 
scenario. 

Searching for {\em direct} CP violation in 
Cabibbo suppressed $D$ decays as a sign for New Physics would also 
represent a very complex challenge: within the KM description one expects 
to find some asymmetries of order 0.1 \%; yet it would be hard 
to conclusively rule out some more or less accidental enhancement due to a 
resonance etc. raising an asymmetry to the 1\% level. 

The only clean environment is provided by CP violation involving 
$D^0$ oscillations, like in $D^0(t) \to K^+ K^-$ and/or 
$D^0(t) \to K^+ \pi ^-$. For the asymmetry would depend 
on the product sin$(\Delta m_D t) \cdot {\rm Im}
[T(\bar D\to f)/T(D\to \bar f)]$: with both factors being 
$\sim
{\cal O}(10^{-3})$ in the SM one predicts a practically zero 
effect.

%%%%%%%%%%%%
\subsection{$P_{\perp}(\mu )$ in $K^+ \to \mu ^+ \pi ^0 \nu$}
%%%%%%%%%%%%%%

The muon polarization transverse to the decay plane in 
$K^+ \to \mu ^+ \pi ^0 \nu$ represents a T-odd correlation 
$P_{\perp}(\mu ) = \langle \vec s(\mu ) \cdot 
(\vec p (\mu) \times \vec p (\pi))/
|\vec p (\mu) \times \vec p (\pi)| \rangle$, 
which 
in this case could not be faked realistically 
by final-state interactions and
would  reveal genuine T violation. 
With $P_{\perp}(\mu ) \sim 10^{-6}$ in the SM, it would also 
reveal New Physics that has to involve chirality breaking weak 
couplings: $P_{\perp}(\mu ) \propto {\rm Im}\xi$, where 
$\xi \equiv f_-/f_+$ with 
$f_-[f_+]$ denoting the chirality violating [conserving] 
decay amplitude. There are `ancient' data yielding 
\beq 
{\rm Im} \xi = -0.01 \pm 0.019 \; \leftrightarrow \; 
P_{\perp}(\mu ) = (-1.85 \pm 3.6)\cdot 10^{-3} 
\eeq 
A new preliminary result from the ongoing experiment 
was reported here: 
\beq 
{\rm Im} \xi = - 0.013 \pm 0.016 \pm 0.003 \; . 
\eeq  

%%%%%%%%%%
\subsection{EDM's}
%%%%%%%%%%%
Electric dipole moments $d$ 
of non-degenerate systems represent direct 
evidence for T violation. The present bounds read: 
\bea 
d_{neutron} &<& 9.7 \cdot 10^{-26} \; ecm \\ 
d_{electron} &=& (-0.3 \pm 0.8) \cdot 10^{-26} \; ecm
\eea
With the KM scheme predicting unobservably tiny effects 
(with the only exception being the `strong CP' problem), 
and many New Physics scenarios yielding 
$d_{neutron}$, $d_{electron}$ $\geq 10^{-27}$ ecm, this is truly a 
promising zero background search for New Physics! 

%%%%%%%%%%%%%
\subsection{KM Trigonometry}
%%%%%%%%%%%%

There certainly exists the potential for a `qualitative' discrepancy 
in the CP asymmetries for $B$ decays. The cleanest case is given by 
the CP asymmetry in $B_s(t) \to \psi \eta$ or 
$B_s(t) \to \psi \phi$, which is Cabibbo suppressed 
\cite{BS80} and thus below 
4\% due to three-family unitarity. 

Yet otherwise the situation in $\Delta B = 1,2$ is more 
complex meaning it provides more opportunites, yet also more 
challenges. For one will be looking for {\em quantitative} 
discrepancies between predictions and the data that 
can{\em not} amount to 
orders of magnitude. 

With three families there are actually six unitarity triangles. They contain 
three types of angles: 
\begin{enumerate}
\item 
Angles of order unity like $\phi_{1,2,3}$; they differ from each other 
in order $\lambda ^2$. 
\item 
Angles that themselves are of order $\lambda ^2$; the most accessible 
representative is an angle in the $bs$ triangle often referred to 
as $\chi$: 
\beq 
\chi = \phi _1^{bs} = \pi + {\rm arg}\left( 
\frac{V_{cs}^*V_{cb}}{V_{ts}^*V_{tb}}\right) \simeq \lambda ^2\eta
\eeq 
which controls the aforementioned asymmetry in 
$B_s(t) \to \psi \phi , \, \psi \eta$ \cite{BS80}.  
\item 
Angles $\sim {\cal O}(\lambda ^4)$, the least unaccessible one 
being in the $cu$ triangle often referred to as $\chi ^{\prime}$ 
\beq 
\chi ^{\prime} = \phi _3^{cu} = {\rm arg}\left( 
\frac{-V_{ud}^*V_{cd}}{V_{us}^*V_{cs}}\right) \simeq 
- \lambda ^4 A^2\eta \; ; 
\eeq
it controls CP asymmetries in $D$ decays \cite{TAUCHARM}. 

\end{enumerate} 
A comprehensive program will have to undertake three steps: 
\begin{itemize}
\item 
measure the large angles $\phi _{1,2,3}$ (and their 
`cousins') and check their correlations with the sides of the 
triangle; 
\item 
check whether the small [tiny] angle $\chi$ [$\chi ^{\prime}$] is  
indeed small [tiny]; 
\item 
attempt to measure the ${\cal O}(\lambda ^2)$ differences 
between $\phi _{1,2,3}$ and their cousins.   

\end{itemize} 
All of these represent searches for New Physics with in particular the 
last item probing features of such New Physics beyond its `mere' 
existence. 

With many of the SM effects being large or at least sizeable, one 
is looking for deviations from expectations that are mostly 
of order unity. A typical scenario would be 
that an asymmetry of, 
say, 40 \% is expected, yet 80\% is observed; how confident 
could we be in claiming New Physics? What about 40\% vs. 
60\% or even 50\%? The situation is thus qualitatively 
different from $K$ decays where {\em original} expectations and data 
differed by orders of magnitude! Therefore we have to be 
very conscious of three 
scourges: (i) Systematic experimental uncertainties; 
(ii) experiments could be wrong -- an issue addressed by the 
`combiner' program \cite{COMBINER}; (iii) theoretical 
uncertainties! 

%%%%%%%%%%%%%%%%%%
\subsection{On Theoretical Uncertainties}
%%%%%%%%%%%%%%%

While considerable experience and awareness exists concerning 
the {\em quantitative} aspects of experimental shortcomings, this is 
not so with respect to theoretical uncertainties. 
My understanding behind
quoting the latter is the  following: "I would be very surprised if the 
true value would fall outside the stated range." Such a 
statement is obviously hard to quantify. 

An extensive literature on how to evaluate them has emerged over the last 
two years in particular 
(see, for example, \cite{STOCCHI,COMBINER}). 
It seems to me that the passion of
the debate  has overshadowed the fact that a lot of learning has happened. For 
example it is increasingly understood that any value within a stated 
range has to be viewed as equally likely. While concerns are 
legitimate that some actors might be overly 
aggressive in stating constraints on the KM triangle, it would be 
unfair to characterize them as silly. I also view it 
as counterproductive to bless one approach while anathematizing all others 
`ex cathedra'. I believe many different paths should be pursued 
since "good decisions come from experience that often is learnt from  
bad decisions". 

Our most powerful weapon for controlling theoretical uncertainties 
will again be  
{\em over}determining basic quantities by extracting their values from more 
than one independant measurement. In this respect the situation is 
actually more favourable in $B$ than in $K$ decays since there 
are fewer free parameters {\em relative} to the number of available 
decay modes. Once the investment has been made to collect the huge 
number of decays required 
to obtain a sufficient number of the transitions  
of primary interest -- say $B_d \to \psi K_S \to (l^+l^-)_{\psi} 
(\pi ^+\pi ^-)_{K_S}$ -- then we have also a slew of many other 
channels that can act as cross checks or provide us with information 
about hadronization effects etc.  Finally one should clearly distinguish 
the goal one has in mind: does one want to state the most likely 
expectation -- or does one want to infer the presence of New Physics 
from a
discrepancy between  expectations and data? The latter goal is of 
course much more ambitious where for once being conservative is a virtue!

%%%%%%%%%%%%%%
\subsection{Looking into the Crystal Ball}
%%%%%%%%%%%%%

I expect various large CP asymmetries to 
be found in $B$ decays -- including direct CP violation -- 
over the next 15
years that agree with the KM expectations to first order, yet exhibit  
smallish, though definite deviations thus revealing the 
intervention of New Physics. However it is conceivable that the 
whole future beauty phenomenology can be accommodated in the 
CKM ansatz. Would that mean our efforts will have been wasted? 

My answer is an emphatic no! The pattern in the Yukawa couplings 
often referred to as `textures' is presumably determined by 
very high scale dynamics. They provide the seeds for the quark 
mass matrix arising when Higgs fields develop vacuum expectation 
values at much lower scales. The quark mass matrix yields the 
quark masses and the CKM angles and phase. My conjecture is that 
such textures follow a simple pattern yielding `special' 
CKM parameters. From the observed values of 
CKM quantities one can thus 
infer information on the dynamics at very 
high scales. 

Yet what is a manifestly simple 
pattern at very high energies will look quite different at 
the electroweak scales that can be probed: renormalization 
will tend to wash out striking features. This again calls for 
{\em precise} extractions of these fundamental parameters.    

%%%%%%%%%%%%
\section{Conclusions \& Outlook}
%%%%%%%%%%%

We have reached an exciting and even decisive phase in 
flavour dynamics. 
\begin{itemize}
\item 
Since the phenomenological success of the CKM description is a 
priori quite surprising, it must contain a deep, albeit  
hidden message. 
\item 
New (sub)paradigms have been established or are about to be 
established: direct CP violation has been found, intriguing 
hints for the first CP asymmetry outside $K_L$ decays have  
emerged and the CKM predictions for CP violation in $B$ decays 
are about to be tested. These represent {\em high sensitivity}  
probes of dynamics and contain many possible portals to 
New Physics. 
\item 
Basic quantities have become known with good accuracy and the 
promise for even more: the beauty quark mass is known to within 
about 1.5 \% -- the most precise quark mass; the top mass is 
known to within 3\% [10 \%] due to 
direct observation [radiative corrections]; $|V_{cb}|$ has been 
extracted with about 5 \% or so accuracy with a reduction 
down to $\sim 2$ \% appearing feasible; the error on 
$|V_{ub}|$ of presently about 40 \% should be reduced to the 
10\% level with 5\% not appearing to be impossible in the long run; 
for $|V_{td}|$ with its present uncertainty $\sim 60$ \% a reduction 
down to 10 \% again might not be impossible. Thus $B$ physics will 
develop into a {\em high precision} probe for New Pghysics as well. 
\item 
These developments have been made possible by 
{\em practical} theoretical technologies having been greatly improved: 
there has been increasing sophistication in treating semileptonic 
and radiative $B$ decays; a new frontier has emerged in treating 
exclusive nonleptonic $B$ decays   
with intriguing classification schemes truly based on 
QCD that might allow us to calculate these transitions in the 
real world. 
\item 
Theoretical uncertainties constitute mostly systematic uncertainties 
with hidden correlations. They can reliably be evaluated only 
through overconstraints. Prior to that they should be 
considered {\em preliminary}; in that context 
I would like to appeal to the 
community to accord us theorists the same professional 
courtesies that is granted to experimental analyses. 
\item 
To make good use of such developments we need experimental programs 
that allow {\em precise} measurements in a {\em comprehensive} 
way rather than just one or two precise ones. It will be an 
exciting adventure to find out how far such a program can be 
pushed. In this context I applaud the managements of CERN 
and FNAL for their wisdom in approving LHC-b and BTeV. 
\item 
There are other areas that might well contain portals to 
New Physics: dedicated searches for CP violation in charm decays, 
EDMs and transverse muon polarization 
in $K_{\mu 3}$ decays are an 
absolute must since any improvement in experimental sensitivity might 
reveal an effect. This is even more so in light of recent efforts 
to explain baryogenesis as being driven by leptogenesis in the 
Universe. 
\item 
We have heard of mounting evidence for neutrino oscillations, which 
require neutrino masses to be nondegenerate implying lepton flavour 
eigenstates to differ from lepton mass eigenstates; the saw-see 
mechanism provides an attractive framework for explaining the 
smallness of neutrino masses. There are intriguing 
connections between the atmospheric neutrino anomaly and 
$\tau \to \mu \gamma$ and between the solar neutrino anomaly 
and $\mu \to e \gamma$ in the context of SUSY GUTs \cite{OKADA}. 

In future meeting there will be detailed 
discussions of the lepton analogue to the Cabibbo-Kobayashi-Maskawa 
matrix, the Maki-Nakagawa-Sakata \cite{MNS} matrix indicating that 
leptons after all are `exactly like quarks -- only different!'.

\end{itemize}

\vskip 3mm  
{\bf Acknowledgements} 
  
The organizers deserve thanks for their smooth organization of this 
conference in the lively city of Osaka. I have benefitted from 
discussions with A. Golutvin, M. Beneke and A. Sanda. 
This work has been supported by the NSF under the grant 
PHY 96-05080. 

%%%%%%%%%%%%%%%%%%%%%%%%%%%%%%%

\end{document}